\documentclass[twoside]{autart}
\usepackage{graphics} 
\usepackage{epsfig} 
\usepackage{times} 
\usepackage{amsmath} 
\usepackage{amssymb}  
\usepackage{float}
\usepackage{epstopdf}
\usepackage{stackengine}
\usepackage{cite}
\usepackage{algpseudocode}
\usepackage{algorithm}
\usepackage{psfrag}
\usepackage{soul}
\usepackage{inputenc}
\usepackage{graphicx}
\usepackage{caption}
\usepackage{subcaption,  balance}
\usepackage{physics, tikz}

\newcommand*\circled[1]{\tikz[baseline=(char.base)]{
\node[shape=circle,draw,inner sep=1.2pt] (char) {#1};}}

\newtheorem{assumption}{Assumption}

\newtheorem{remark}{Remark}

\newtheorem{definition}{Definition}

\newtheorem{proposition}{Proposition}

\begin{document}

\begin{frontmatter}
\vspace{-0.20in}
\runtitle{Orbital Stabilization of Underactuated Systems Using VHC and ICPM}  

\title{Orbital Stabilization of Underactuated Systems using Virtual Holonomic Constraints and Impulse Controlled Poincar\'e Maps\thanksref{footnoteinfo}} 
\thanks[footnoteinfo]{
The authors acknowledge the support provided by the National Science Foundation, Grant CMMI-1462118.\\Corresponding author Ranjan~Mukherjee.}
\vspace{-0.15in}
\author[ME]{Nilay Kant}\ead{kantnila@egr.msu.edu},    
\author[ME]{Ranjan Mukherjee}\ead{mukherji@egr.msu.edu}         

\address[ME]{Department of Mechanical Engineering, Michigan State University, East Lansing, USA}                                     

\begin{keyword}                           
Impulsive control, orbital stabilization, Poincar\'e map, underactuated system, virtual holonomic constraint           
\end{keyword}                             

\begin{abstract}                          
The problem of orbital stabilization of underactuated mechanical systems with one passive degree-of-freedom (DOF) is revisited. Virtual holonomic constraints are enforced using a continuous controller; this results in a dense set of closed orbits on a constraint manifold. A desired orbit is selected on the manifold and a Poincar\'e section is constructed at a fixed point on the orbit. The corresponding Poincar\'e map is linearized about the fixed point; this results in a discrete linear time-invariant system. To stabilize the desired orbit, impulsive inputs are applied when the system trajectory crosses the Poincar\'e section; these inputs can be designed using standard techniques such as LQR. The Impulse Controlled Poincar\'e Map (ICPM) based control design has lower complexity and computational cost than control designs proposed earlier. The generality of the ICPM approach is demonstrated using the 2-DOF cart-pendulum and the 3-DOF tiptoebot.
\end{abstract}
\vspace{-0.10in}
\end{frontmatter}

\section{Introduction} \label{sec1}
For underactuated systems, Virtual Holonomic Constraint (VHC) based control designs have gained popularity due to their conceptual simplicity and applicability to control of repetitive motion; they have been used for gait stabilization in bipeds \cite{grizzle2014models, westervelt2018feedback, grizzle2001asymptotically, canudas2004concept } and trajectory control for systems with open kinematic chains \cite{canudas2002orbital, mohammadi2018dynamic, maggiore2012virtual, shiriaev2005constructive, shiriaev2007virtual, freidovich2008periodic, mohammadi2015maneuvering, shiriaev2006periodic, shiriaev2010transverse}. VHCs parameterize the active joint variables in terms of the passive joint variables and confine system trajectories to a constraint manifold \cite{maggiore2012virtual}. To enforce the VHC, the constraint manifold has to be stabilized using feedback. Typically, a constraint manifold contains a dense set of periodic orbits and the choice of repetitive motion determines the specific orbit that has to be stabilized. To stabilize biped gaits, for example, Grizzle et.al \cite{grizzle2001asymptotically, westervelt2018feedback} enforced the VHC and periodic loss of energy due to ground-foot interaction was exploited for orbital stabilization.\

A special class of underactuated systems are those with one passive DOF. For such systems, Shiriaev and collaborators \cite{shiriaev2005constructive, shiriaev2007virtual, freidovich2008periodic} used VHC to select the desired orbit. For an $n$-DOF system, the $2n$ dimensional dynamics is linearized about the desired orbit; this results in a ${2n\!-\!1}$ dimensional system. A periodic Ricatti equation is then solved to design a time-varying controller that stabilizes the orbit. It should be noted that the control designs in \cite{shiriaev2005constructive, shiriaev2007virtual, freidovich2008periodic} stabilize the orbit but do not enforce the VHC. A control scheme that enforces the VHC and simultaneously stabilizes the orbit was recently proposed in \cite{mohammadi2018dynamic}. The key idea is that the VHC is made time-varying using a scalar parameter which is controlled via feedback. The stabilization problem involves solving a periodic Ricatti equation; however, unlike \cite{shiriaev2005constructive, shiriaev2007virtual, freidovich2008periodic}, where the dimension of the system is ${2n\!-\!1}$, the dimension of the system in \cite{mohammadi2018dynamic} is always three. For systems with more than two DOF, the method in \cite{mohammadi2018dynamic} reduces the computational complexity of control implementation. Also, by enforcing the VHC, it improves control over transient characteristics of the trajectory \cite{mohammadi2016virtual}. Similar to \cite{mohammadi2016virtual, mohammadi2018dynamic}, we propose a control design that enforces the VHC and stabilizes the desired orbit. The control design is comprised of continuous inputs that enforce the VHC and impulsive inputs that exponentially stabilize the orbit. Impulsive inputs have been used for control of underactuated systems \cite{mathis2014impulsive, jafari2016enlarging, kant2018impulsive, kant2019estimation, kant2019, albahkali2009swing} and it has been established that such inputs can be implemented in standard hardware using high-gain feedback.

This paper is organized as follows. The system dynamics is presented in section \ref{sec2} and the results in \cite{maggiore2012virtual} are utilized to enforce VHC such that the resulting zero dynamics is Euler-Lagrange. A periodic orbit is selected on the constraint manifold and a method for orbital stabilization is presented in section \ref{sec4}. To stabilize the orbit, a Poincar\'e section is defined at a point on the orbit and the return map is linearized about the fixed point; this results in a ${2n\!-\!1}$ dimensional discrete linear time-invariant (LTI) system. To control this system and stabilize the orbit, impulsive inputs are applied when the system trajectory crosses the Poincar\'e section. The controllability of the orbit can be verified by simply checking the controllability of the linear system. This is simpler than the approach in \cite{shiriaev2005constructive, shiriaev2007virtual, freidovich2008periodic} where controllability is verified numerically along the orbit for most systems. Since the system is LTI, the control design involves constant gains that can be computed off-line. Compared to the methods proposed earlier \cite{shiriaev2005constructive, shiriaev2007virtual, freidovich2008periodic, mohammadi2018dynamic}, where periodic Ricatti equations have to be solved, our method has lower computational cost and complexity. Since impulsive inputs are used to control the Poincar\'e map, the closed-loop system dynamics can be described by the Impulse Controlled Poincar\'e Map (ICPM). The simplicity and generality of the ICPM approach to orbital stabilization is demonstrated using the examples of the 2-DOF cart-pendulum in section \ref{sec-6} and the 3-DOF tiptoebot in section \ref{sec-5}. Concluding remarks are presented in section \ref{sec7}.\

\section{Problem Formulation}\label{sec2}
\subsection{System Dynamics}\label{sec2-1}
Consider an $n$ DOF underactuated system with one passive DOF, where the passive DOF is a revolute joint. Let $q$, $q \triangleq \left[q_1^T \,\, q_2\right]^T$, denote the generalized coordinates, where $q_1 \in R^{n-1}$ and $q_2 \in S$, $S = R$ modulo $2\pi$, are the coordinates of the active and the passive DOFs. The configuration space of the system is denoted by $\mathbb{Q}^n$, $\mathbb{Q}^n \in R^{n-1} \times S$. The Lagrangian of the system can be written as
\begin{align*}
L(q, \dot q) = \frac{1}{2}\dot q^T M(q) \,\dot q + \mathcal{F}(q)
\end{align*}
\noindent In the equation above, $M(q) \in R^{n\times n}$ denotes the symmetric, positive-definite mass matrix, partitioned as
\begin{align*}
M(q) &= \left[\begin{array}{c|c} M_{11}(q) &M_{12}(q)\\[0.25ex] \hline M_{12}^T(q) &M_{22}(q)\end{array}\right]
\end{align*}
\noindent where $M_{11} \in R^{(n-1) \times (n-1)}$, $M_{22} \in R$ and $\mathcal{F}(q)$ is the potential energy of the system. The Euler-Lagrange equation of motion can be written as follows
\begin{subequations} \label{eq1}
\begin{align}
M_{11}(q)\,\ddot q_1 + M_{12}(q)\,\ddot q_2 + h_1 (q, \dot q) &= u\label{eq1a}\\
M_{12}^T(q)\,\ddot q_1 + M_{22}(q)\,\ddot q_2 + h_2(q, \dot q) &= 0\label{eq1b}
\end{align}
\end{subequations}
\noindent where $u \in R^{n-1}$ is the control input, and $[h_1^T, h_2]^T$ is the vector of Coriolis, centrifugal and gravity forces. In compact form, (\ref{eq1a}) and (\ref{eq1b}) can be rewritten as
\begin{subequations} \label{eq2}
\begin{align}
\ddot q_1 &= A(q, \dot q) + B(q) u \label{eq2a} \\
\ddot  q_2 &= C(q, \dot q) +D(q) u \label{eq2b}
\end{align}
\end{subequations}
\noindent where,
\begin{equation} \label{eq3}
\begin{aligned}
B(q) &= \left[M_{11} - (1/M_{22}) M_{12}\,M_{12}^{T}\right]^{-1} \cr
A(q, \dot q) &= (1/M_{22}) B(q) \left[M_{12}\,h_2 - h_1 M_{22}\right] \cr
D(q) &=-(1/M_{22})M^T_{12}\,B(q) \cr
C(q, \dot q) &= -(1/M_{22})\left[M^T_{12}\, A(q, \dot q) + h_2\right]
\end{aligned}
\end{equation}
\noindent Similar to \cite{maggiore2012virtual}, we make the following assumption:
\begin{assumption}\label{assum-1}
For some $\bar q \triangleq [\bar q_1^T, \bar q_2]^T \in \mathbb{Q}^n$, the mass matrix $M(q)$ and the potential energy $\mathcal{F}(q)$ are even with respect to $\bar q$, \emph{i.e.},
$$M(\bar q + q) = M(\bar q -q), \quad \mathcal{F}(\bar q + q) = \mathcal{F}(\bar q -q)$$
\end{assumption}
\vspace{-0.20in}

\subsection{Imposing Virtual Holonomic Constraints (VHC)}
A holonomic constraint enforced by feedback is referred to as VHC. The current and the next subsection summarizes relevant results from \cite{maggiore2012virtual}. For a wide class of mechanical systems, a comprehensive discussion on VHC can be found in \cite{maggiore2012virtual}, \cite{mohammadi2016virtual}.\

A VHC for (\ref{eq1}) is described by the relation $\rho(q) = 0$ where, $\rho : \mathbb{Q}^n \rightarrow R^{n-1}$ is smooth and rank$[J_{q}(\rho)] = {n\!-\!1}$ for all $q \in \rho^{-1}(0)$. Here, $J_q(\rho)$ is the Jacobian of $\rho$ with respect to $q$. The VHC is said to be stabilizable if there exists a smooth  feedback $u_c(q, \dot q)$ that asymptotically stabilizes the set
\begin{align}\label{eq4}
\mathcal{C} &= \{ (q, \dot q) : \rho(q) =0,\,\, J_{q}(\rho)\dot q = 0 \}
\end{align}
\noindent The set $\mathcal{C}$, which is referred to as  the constraint manifold,  is controlled invariant \cite{maggiore2012virtual}. For the system described by (\ref{eq1}), $\mathcal{C}$ is an ${(n\!-\!1)}$ dimensional manifold.  \

An important goal of this paper is to generate repetitive motion, which can be described by closed orbits. Consequently, $\rho^{-1}(0)$ must be a smooth and closed curve without any self-intersection. The VHC can be described as
\begin{equation}\label{eq7}
\rho(q) = q_1 - \Phi(q_2) = 0
\end{equation}
%
\noindent where $\Phi : S \rightarrow R^{n-1}$ is a smooth vector-valued function.
\noindent The constraint manifold $\mathcal{C}$ in (\ref{eq4}) can be expressed as:
\begin{align}\label{eq8}
\mathcal{C} &= \bigg\{ (q, \dot q) :q_1 = \Phi(q_2),\,\, \dot q_1 = \left[\frac{\partial \Phi}{\partial q_2}\right]\dot q_2 \bigg\}
\end{align}
\noindent It should be noted that since $q_2 \in S$, $\Phi(q_2 + 2\pi) = \Phi(q_2)$ and $\rho^{-1}(0)$ is closed. Following the notion of odd VHC \cite{maggiore2012virtual}, we state another assumption.
\begin{assumption}\label{assum-2}
For $\bar q$ which satisfies Assumption \ref{assum-1}, $\Phi(q_2)$ is odd with respect to $\bar q_2$, \emph{i.e.},
$$\Phi(\bar q_2 + q_2) = -\Phi(\bar q_2 - q_2)$$
\end{assumption}
\vspace{-0.10in}
To stabilize $\mathcal{C}$, we investigate the dynamics of $\rho(q)$; differentiating $\rho(q)$ twice with respect to time, we get
\begin{align}\label{eq9}
\ddot \rho = \ddot q_1 - \left[\frac{\partial \Phi}{\partial q_2}\right]\ddot q_2 - \left[\frac{\partial^2 \Phi}{\partial q_2^2}\right]\dot q_2^2
\end{align}
\noindent Substitution of $\ddot q_1$ and $\ddot q_2$ from (\ref{eq2a}) and (\ref{eq2b}) in (\ref{eq9}) yields
\begin{align}\label{eq10}
\ddot \rho = A - \left[\frac{\partial^2 \Phi}{\partial q_2^2}\right]\dot q_2^2 - \left[\frac{\partial \Phi}{\partial q_2}\right]C+ \left[B - \left[\frac{\partial \Phi}{\partial q_2}\right]D\right]u
\end{align}
\noindent The following choice of linearizing control
\begin{equation}
\begin{aligned}\label{eq11}
u_c = \left[B -\left[\frac{\partial \Phi}{\partial q_2}\right]D\right]^{-1} &\left[-A + \left[\frac{\partial^2 \Phi}{\partial q_2^2}\right]\dot q_2^2 \right. \cr
+ &\left. \left[\frac{\partial \Phi}{\partial q_2}\right]C - k_p\rho - k_d \dot \rho\right]
\end{aligned}
\end{equation}
\noindent where $k_p$ and $k_d$ are positive definite matrices, results in
\begin{align}\label{eq12}
\ddot \rho + k_d \dot \rho + k_p \,\rho =0
\end{align}
\noindent This implies that $\lim_{t \rightarrow \infty} \rho(t) \rightarrow 0$ exponentially and $u_c$ in (\ref{eq11}) stabilizes the VHC in (\ref{eq7}). If the initial conditions are chosen such that $\rho(0) = \dot \rho(0) =0$, $u_c$ in (\ref{eq11}) enforces the VHC and the constraint manifold $\mathcal{C}$ is controlled invariant.\
\begin{remark}\label{rem1}
For $u_c$ in (\ref{eq11}) to be well-defined, the matrix $\left[B - ({\partial \Phi}/{\partial q_2})D\right]$ must be invertible. It can be shown that 
$\left[B - ({\partial \Phi}/{\partial q_2})D\right]$ is invertible iff $M^T_{12}({\partial \Phi}/{\partial q_2}) + M_{22} \neq 0$. This is also a necessary and sufficient condition for $\mathcal{C}$ to be stabilizable - see proposition 3.2 of \cite{maggiore2012virtual}.
\end{remark}

\subsection{Zero Dynamics and Periodic Orbits}
On the constraint manifold $\mathcal{C}$, the dynamics of the system satisfies $\rho(q) \equiv 0$; this implies
\begin{align}\label{eq13}
q_1 = \Phi(q_2),\,\, \dot q_1 = \left[\frac{\partial \Phi}{\partial q_2}\right]\dot q_2, \,\, \ddot q_1 =  \left[\frac{\partial^2 \Phi}{\partial q_2^2}\right]\dot q_2^2 + \left[\frac{\partial \Phi}{\partial q_2}\right]\ddot q_2
\end{align}
\noindent Substitution of $q_1$, $\dot q_1$ and $\ddot q_1$ from (\ref{eq13}) in (\ref{eq1b}) provides the zero dynamics, which can be expressed in the following form
\begin{equation}\label{eq14}
\ddot q_2 = \alpha_1(q_2) + \alpha_2(q_2) \dot q_2^2
\end{equation}
\noindent It was shown in \cite{shiriaev2005constructive, shiriaev2006periodic, shiriaev2010transverse, maggiore2012virtual} that the equation above has an integral of motion of the form
\begin{equation}\label{eq15}
\begin{aligned}
E(q_2, \dot q_2) &= (1/2)\mathcal{M}(q_2)\dot q_2^2 + \mathcal{P}(q_2) \cr
\mathcal{M}(q_2) &= {\rm{exp}} \left(-2\int_{0}^{q_2} \alpha_2(\tau) d\tau\right)\cr
\mathcal{P}(q_2) &= -\int_{0}^{q_2}\alpha_1(\tau) \mathcal{M}(\tau) \,d\tau
\end{aligned}
\end{equation}
\noindent where $\mathcal{M}(q_2)$ is the mass and $\mathcal{P}(q_2)$ is the potential energy of the reduced system in (\ref{eq14}). Since Assumption \ref{assum-1} and \ref{assum-2} are satisfied, the zero dynamics represents an Euler-Lagrange system with the Lagrangian\footnote{Assumptions \ref{assum-1} and \ref{assum-2} provide necessary and sufficient conditions for the reduced system to be Euler-Lagrange - the proof of this result can be found in\cite{ maggiore2012virtual, mohammadi2016virtual}.} equal to $(1/2)\mathcal{M}(q_2)\dot q_2^2 - \mathcal{P}(q_2)$.

The zero dynamics in (\ref{eq14}) is similar to the dynamics of a simple pendulum and its qualitative properties can be described by the potential energy $\mathcal{P}(q_2)$. Let $\mathcal{P}_{\rm min}$ and
$\mathcal{P}_{\rm max}$ denote the minimum and maximum values of $\mathcal{P}$. If an energy level set is denoted by $E(q_2, \dot q_2) =c$, then $c \in (\mathcal{P}_{\rm min},
\mathcal{P}_{\rm max})$ corresponds to a periodic orbit where the sign of $\dot q_2$ changes periodically and $c > \mathcal{P}_{\rm max}$ corresponds to an orbit where the sign of $\dot q_2$ does not change \cite{maggiore2012virtual}.\

\subsection{Problem Statement}\label{sec3-4}
Since the zero dynamics in (\ref{eq14}) has an Euler-Lagrange structure, there cannot exist any non-trivial isolated periodic orbit - this follows from the Poincar\'e-Lyapunov-Liouville-Arnol'd theorem \cite{nekhoroshev1994poincare, mohammadi2016virtual, maggiore2012virtual}. A direct implication of this theorem is that the reduced dynamics possesses a dense set of closed orbits, that are unstable. For a desired repetitive motion, the corresponding orbit must be stabilized. Consider the desired closed orbit $\mathcal{O}_d$, defined as follows:
\begin{align}\label{eq16}
\mathcal{O}_d = \{q, \dot q \in \mathcal{C} : E(q_2, \dot q_2) = c_d\}, \quad c_d > \mathcal{P}_{\rm min}
\end{align}
\noindent Let $x$, $x \triangleq [q^T, \dot q^T]^T$, denote the states of the system in (\ref{eq1}). We define an $\epsilon$-neighborhood of $\mathcal{O}_d$ by
\begin{align*}
&U_{\epsilon} = \{x \in \mathbb{Q}^n \times R^n : {\rm{dist}}(x, \mathcal{O}_d) < \epsilon\}\cr
&{\rm{dist}}(x, \mathcal{O}_d) \triangleq \inf_{y\in \mathcal{O}_d} \| x - y\|
\end{align*}
\noindent We now define stability of the orbit $\mathcal{O}_d$ from \cite{khalil2002nonlinear}.
\begin{definition}\label{def1}
The orbit $\mathcal{O}_d$ in (\ref{eq16}) is
\begin{itemize}
\item stable, if for every $\epsilon >0$, there is a $\delta >0$ such that $x(0) \in U_{\delta} \implies x(t) \in U_{\epsilon}, \,\, \forall t \geq 0$.
\item asymptotically stable if it is stable and $\delta$ can be chosen such that $\lim_{t \to \infty} {\rm{dist}}(x(t), \mathcal{O}_d) = 0$.
\end{itemize}
\end{definition}
The control $u_c$ in (\ref{eq11}) stabilizes $\mathcal{C}$ but does not stabilize $\mathcal{O}_d$. If $q, \dot q \in \mathcal{O}_d$, $u_c$ enforces the VHC and trajectories stay on $\mathcal{O}_d$; however, a perturbation of the states will cause the trajectories to converge to a different orbit on $\mathcal{C}$. The objective of this paper is to enforce the VHC and exponentially stabilize the desired orbit $\mathcal{O}_d$ on $\mathcal{C}$.

\section{Main Result: Stabilization of $\mathcal{O}_d$}\label{sec4}
\subsection{Poincar\'e Map}\label{sec4-1}
The system in (\ref{eq1}) with $u = u_c$ defined in (\ref{eq11}), has the state-space representation
\begin{align}\label{eq17}
\dot x= f(x)
\end{align}
\noindent The stability characteristics of periodic orbits can be studied using Poincar\'e maps \cite{strogatz2018nonlinear}. To this end, we define the Poincar\'e section $\Sigma$ of $\mathcal{O}_d$ as follows\footnote{In the definition of $\Sigma$ in (\ref{eq18}), $\dot q_2 \geq 0$ can be replaced with  $\dot q_2 \leq 0$ without any loss of generality.}:
\begin{align}\label{eq18}
\Sigma = \{x \in \mathbb{Q}^n \times R^n \,:\, q_2 = q_2^*, \dot q_2 \geq 0\}
\end{align}
\noindent where $q_2^*$ is a constant. Let $z$, $z \triangleq [q_1^T, \dot q^T]^T \in R^{(2n-1)}$, denote the states of the system on $\Sigma$. The Poincar\'e map $\mathbb{P} : \Sigma \rightarrow \Sigma$ is obtained by following trajectories of $z$ from one intersection with $\Sigma$ to the next. Let $t_k$, $k = 1, 2, \cdots$ denote the time of the $k$-th intersection and $z(k) = z(t_k)$. Then, $z(k+1)$ can be described with the help of the map $\mathbb{P}$
\begin{align}\label{eq19}
z(k+1) &= \mathbb{P}[z(k)]
\end{align}
\noindent The point of intersection of $\Sigma$ and $\mathcal{O}_d$ is the fixed point of $\mathbb{P}$ denoted by $z^*$; it satisfies the following relation
\begin{align}\label{eq20}
z^* = \mathbb{P}(z^*)
\end{align}
\noindent The stability characteristics of the orbit $\mathcal{O}_d$ can be studied by investigating the stability properties of $z^*$, which is an equilibrium point of the discrete-time system in (\ref{eq19}); this can be done by linearizing the map $\mathbb{P}$ about $z^*$. For $z(k) = z^* + \nu$, where $\| \nu \|$ is a small number, we can write
\begin{align}\label{eq21}
z(k+1) &= \mathbb{P}(z^* + \nu) \\
&=\mathbb{P}(z^*) +\left[ \grad_z\mathbb{P}(z)\right]_{z=z^*}[z(k)-z^*] + O(\|\nu\|^2) \notag
\end{align}
\noindent Using $\mathbb{P}(z^*) = z^*$ from (\ref{eq20}) and neglecting higher-order terms in $\| \nu \|$, the above equation can be written as
\begin{align}\label{eq22}
e(k+1) = & \,\mathcal{A}\, e(k)\cr
e(k) \triangleq z(k) - z^*, \quad &\mathcal{A} \triangleq  \left[\grad_z\mathbb{P}(z)\right]_{z=z^*}
\end{align}
\noindent The stability properties of $z^*$ is governed by the eigenvalues of $\mathcal{A}$, which are referred to as the Floquet multipliers of $\mathcal{O}_d$. If the Floquet multipliers lie inside the unit circle, $\mathcal{O}_d$ is exponentially stable - see Theorem 7.3 of \cite{khalil2002nonlinear}. From our discussion in section \ref{sec3-4} we know that the desired orbit $\mathcal{O}_d$ is unstable, \emph{i.e.}, not all eigenvalues of $\mathcal{A}$ lie inside the unit circle. To stabilize the orbit, \emph{i.e.}, to stabilize $z^*$, we design an impulse controller in the next subsection.\

\subsection{Impulse Controlled Poincar\'e Map (ICPM)}\label{sec4-2}
To stabilize the desired orbit $\mathcal{O}_d$, our controller is modified as follows
\begin{equation} \label{eq23}
u = u_c + u_{\mathcal I}
\end{equation}
\noindent where $u_{\mathcal I}$ is an impulsive input which is applied only when $x(t) \in \Sigma$. The dynamics of the system with $u_{\mathcal I}$ as the new input can be written as
\begin{subequations} \label{eq24}
\begin{align}
M_{11}(q)\,\ddot q_1 + M_{12}(q)\,\ddot q_2 + \bar{h}_1 (q, \dot q) &= u_{\mathcal I} \label{eq24a}\\
M_{12}^T(q)\,\ddot q_1 + M_{22}(q)\,\ddot q_2 + h_2(q, \dot q) &= 0\label{eq24b}
\end{align}
\end{subequations}
\noindent where $\bar{h}_1 \triangleq (h_1 - u_c)$. Impulsive inputs cause  discontinuous changes in the generalized velocities while there is no change in the generalized coordinates. On the Poincar\'e section $\Sigma$, the jump in velocities can be computed by integrating (\ref{eq23}) as follows \cite{flynn2010active}:
\begin{equation}\label{eq25}
\left[\!\!\begin{array}{cc} M_{11} &M_{12} \cr M_{12}^T &M_{22}\end{array}\!\!\right] \left[\!\!\begin{array}{c} \Delta \dot q_1 \cr \Delta \dot q_2 \end{array}\!\!\right] = \left[\begin{array}{c} \mathcal{I} \cr 0 \end{array}\right],\quad \mathcal{I} \triangleq \int_0^{\Delta t} u_{\mathcal I}\, {dt}
\end{equation}
\noindent In the above equation, ${\Delta t}$ is the infinitesimal interval of time for which $u_{\mathcal I}$ is active, $\mathcal{I} \in R^{n-1}$ is the impulse of the impulsive input, and $\Delta \dot q_1$ and $\Delta \dot q_2$ are defined as
\begin{equation}\label{eq25a}
\Delta \dot q_1 \triangleq (\dot q_1^{+} - \dot q_1^{-}), \qquad \Delta \dot q_2 \triangleq (\dot q_2^{+} - \dot q_2^{-})
\end{equation}
\noindent where $\dot q^{-}$ and $\dot q^{+}$ are the velocities immediately before and after application of $u_{\mathcal I} $. Since the system is underactuated, the jump in the passive velocity $\dot q_2$ is dependent on the jumps in the active velocity $\dot q_1$; this relationship is described by the {$(n\!-\!1)$} dimensional impulse manifold \cite{jafari2016enlarging, kant2019estimation}, which can be obtained from (\ref{eq25}):\
\begin{equation}\label{eq26}
\mathcal{I}_{\text M} = \{\dot q_1^{+}, \dot q_2^{+} \mid \Delta \dot q_2 = - (1/M_{22}) M_{12}^T \,\Delta \dot q_1 \}
\end{equation}
Since impulsive inputs can cause the system states to move on $\Sigma$, we exploit this property to design a feedback law that stabilizes $z^*$, \emph{i.e.}, stabilizes $\mathcal{O}_d$. The control input applied at $t_k$ is denoted by $\mathcal{I}(k)$\footnote{As long as $\Delta t$ is sufficiently small, the effect of the impulsive input $u_{\mathcal I}$ depends solely on the value of $\mathcal{I}$ - see (\ref{eq25}). Thus $\mathcal{I}$ can be viewed as the control input.}. The dynamics of the impulse controlled system in (\ref{eq19}) can be described by the map
\begin{align}\label{eq27}
z(k+1) &= \mathbb{P}[z(k), \mathcal{I}(k)]
\end{align}
\noindent where $\mathcal{I}(k) = 0$ if $z(k) = z^*$. By linearizing the above map about the fixed point $z^*$ and $\mathcal{I} = 0$, we get
\begin{equation}\label{eq28}
\begin{aligned}
e(k+1) &= \mathcal{A}\,e(k) + \mathcal{B}\,\mathcal{I}(k) \cr
\mathcal{A} &\triangleq \Big[\grad_z\mathbb{P}(z, \mathcal{I})\Big]_{\substack{z=z^*\!\!,\, \mathcal{I}=0}} \cr
\mathcal{B} &\triangleq \Big[\grad_{\mathcal{I}}\mathbb{P}(z, \mathcal{I})\Big]_{\substack{z=z^*\!\!,\, \mathcal{I}=0}}
\end{aligned}
\end{equation}
\noindent where $\mathcal{A} \in R^{(2n-1) \times (2n-1)}$ and $\mathcal{B} \in R^{(2n-1) \times (n-1)}$ can be obtained numerically. Since $\mathcal{A}$ is not Hurwitz (see discussion in the last sub-section), we make the following proposition to stabilize $\mathcal{O}_d$:
\begin{proposition}\label{prop1}
If the pair $\{\mathcal{A}, \mathcal{B}\}$ is stabilizable, the orbit $\mathcal{O}_d$ can be stabilized using the discrete impulsive feedback
\begin{equation}\label{eq29}
\mathcal{I}(k) = \mathcal{K}\, e(k)
\end{equation}
\noindent where the matrix $\mathcal{K}$ is chosen such that $(\mathcal{A} + \mathcal{B}\mathcal{K})$ is Hurwitz.
\end{proposition}
\begin{figure}[t!]
\centering
\psfrag{A}[][]{\small{$\Sigma$}}
\psfrag{B}[][]{\small{$\mathcal{I}_{\text M}$}}
\psfrag{C}[][]{\small{$z^*$}}
\psfrag{D}[][]{\tiny{$\circled{1}$}}
\psfrag{E}[][]{\tiny{$\circled{2}$}}
\psfrag{F}[][]{\tiny{$\circled{3}$}}
\psfrag{G}[][]{\tiny{$\circled{4}$}}
\psfrag{H}[][]{\small{$\mathcal{O}_d$}}
\includegraphics[width=0.6\linewidth]{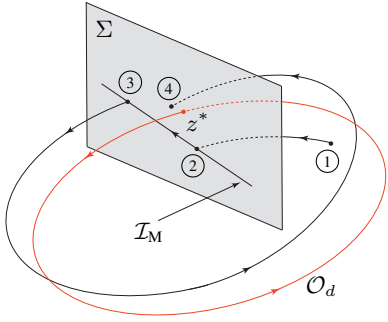}
\caption{Schematic of ICPM approach to orbital stabilization.}  
\label{Fig7}
\end{figure}
The above approach to stabilization, which we refer to as the impulse controlled Poincar\'e map (ICPM) approach, is explained with the help of the schematic in Fig.\ref{Fig7}. The desired orbit $\mathcal{O}_d$ is shown in red and it intersects the Poincar\'e section $\Sigma$ at the fixed point $z^*$. A trajectory starting from an arbitrary initial condition, shown by the point $\circled{\small{1}}$, intersects $\Sigma$ at $\circled{\small {2}}$. The impulsive input in (\ref{eq29}) moves the configuration of the system from $\circled{\small{2}}$ to $\circled{\small{3}}$ along the impulse manifold $\mathcal{I}_{\text M}$, where
$\mathcal{I}_{\text M} \subset \Sigma$. In other words, we impose the restriction that $\circled{\small{3}}$ lies on $\Sigma$, \emph{i.e.}, $\dot q_2^{+} \geq 0$. Hereafter, the altered system trajectory evolves under the continuous control $u_c$ and $\circled{\small{4}}$ denotes its next intersection with $\Sigma$. A series of ICPMs, similar to the map $\circled{\small{2}} \rightarrow \circled{\small{4}}$ exponentially converge the intersection point of the trajectory on $\Sigma$ to $z^*$.

\subsection{Implementation of Control Design}\label{sec4-3}
\subsubsection{Numerical Computation of $\mathcal{A}$ and $\mathcal{B}$ matrices}\label{sec4-3-1}
Let $\delta_i$, $i = 1, 2, \cdots, {2n\!-\!1}$, denote the $i$-th column of $\varepsilon_1 I_{(2n-1)}$, where $\varepsilon_1$ is a small number and $I_{(2n-1)}$ is the identity matrix of size {$(2n\!-\!1)$}. If
$\mathcal{A}_i$ denotes the $i$-th column of $\mathcal{A}$, then $\mathcal{A}_i$ can be numerically computed as follows:
\begin{equation}\label{eq30}
\mathcal{A}_i = \frac{1}{\varepsilon_1} \left[\mathbb{P} (z^* + \delta_i) - z^*\right]
\end{equation}
Let $Q \in R^{n \times (n-1)}$ and $S \in R^{(n-1)}$ be defined as follows:
\begin{equation*}
Q \triangleq \left[\begin{array}{c} I_{(n-1)} \\[0.15ex] \hline 0_{1 \times (n-1)}\end{array}\right], \quad S \triangleq \left[\begin{array}{c} 0_{(n-1) \times 1} \\[0.15ex] \hline M(q)^{-1} \eta_i\end{array}\right]
\end{equation*}
\noindent where $0_{i \times j}$ is a matrix of zeros of dimension $i \times j$, and $\eta_i$, $i = 1, 2, \cdots, {n\!-\!1}$, denote the $i$-th column of $\varepsilon_2 Q$, where $\varepsilon_2$ is a small number. If $\mathcal{B}_i$ denotes the $i$-th column of $\mathcal{B}$, then $\mathcal{B}_i$ can be numerically computed as follows:
\begin{equation}\label{eq31}
\mathcal{B}_i = \frac{1}{\varepsilon_2} \left\{\left[\mathbb{P} (z + S)\right]_{z = z^*} - z^*\right\}
\end{equation}
\noindent The above expression has been obtained using (\ref{eq25}).\

\subsubsection{Impulsive Input using High-Gain Feedback}\label{sec4-3-2}
Impulsive inputs are Dirac-delta functions and cannot be realized in real physical systems. Using singular perturbation theory \cite{kokotovic1999singular}, it was shown that continuous-time implementation of impulsive inputs can be carried out using high-gain feedback \cite{jafari2016enlarging}. To obtain the expression for the high-gain feedback, we substitute (\ref{eq29}) in (\ref{eq25}) to get
\begin{equation}\label{eq32}
\Delta \dot q_1(k) = B\,\mathcal{I}(k) = B\mathcal{K} e(k)
\end{equation}
\noindent where $B$ is defined in (\ref{eq3}) and is evaluated at $t_k$; $\Delta \dot q_1(k)$ is the jump in the active velocities generated by the input $\mathcal{I}(k)$. From (\ref{eq25a}) and (\ref{eq32}), the desired active joint velocities at $t_k$ is
\begin{equation}\label{eq33}
\dot q_1^{\rm des}(k) = \dot q_1(k) + B\mathcal{K} e(k)
\end{equation}
\noindent where $\dot q_1(k) = \dot q_1(t_k)$. To reach the desired velocities in a very short period of time, we use the high-gain feedback \cite{kant2019estimation}
\begin{equation}\label{eq34}
u_{\rm hg} =
B^{-1} \left[\dfrac{1}{\mu}\!\Lambda \left(\dot{q}_1^{\rm des}(k) -  \dot{q}_1 \right) - \bar{A} \right]
\end{equation}
\noindent which remains active for as along as $\| \dot{q}_1^{\rm des}(k) -  \dot{q}_1 \| \geq \varepsilon_3$, where $\varepsilon_3$ is a small number. In (\ref{eq34}), $\dot{q}_1^{\rm des}$ is obtained from (\ref{eq33}) and $\bar A$ is obtained from the expression for $A$ in (\ref{eq3}) by replacing $h_1$ with $\bar h_1$. Furthermore, $\Lambda \triangleq {\rm diag }[\begin{matrix}\lambda_1 &\lambda_2 &\cdots &\lambda_{n-1} \end{matrix}]$, where $\lambda_i$, $i = 1, 2, \cdots, {n\!-\!1}$ are positive numbers, and $\mu > 0$ is a small number.\

\section{Illustrative Example: Cart-Pendulum} \label{sec-6}
\subsection{System Dynamics and VHC} \label{sec-6-1}
\begin{figure}[b!]
\centering
\psfrag{A}[][]{\footnotesize{$x$}}
\psfrag{B}[][]{\footnotesize{$u$}}
\psfrag{C}[][]{\footnotesize{$\theta$}}
\psfrag{D}[][]{\footnotesize{$m_p$}}
\psfrag{E}[][]{\footnotesize{$\ell$}}
\psfrag{F}[][]{\footnotesize{$m_c$}}
\psfrag{M}[][]{\footnotesize{$g$}}
\includegraphics[width=0.60\linewidth]{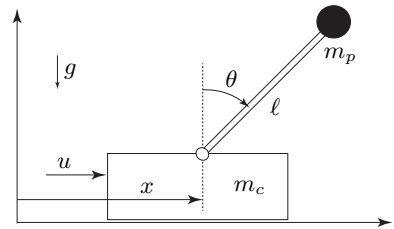}
\caption{Inverted pendulum on a cart.}  
\label{Fig4}
\end{figure}
Consider the frictionless cart-pendulum system in Fig.\ref{Fig4}. The masses of the cart and pendulum are denoted by $m_c$  and $m_p$, $\ell$ denotes the length of the pendulum, and $g$ is the acceleration due to gravity. The control input $u$ is the horizontal force applied on the cart. The cart position is denoted by $x$ and the angular displacement of the pendulum, measured clock-clockwise with respect to the vertical, is denoted by $\theta$. We consider physical parameters of the system to be the same as those in \cite{shiriaev2005constructive}: $m_p = m_c = \ell =1$ . With the following definition
\begin{align}\label{eq35}
q = [\begin{matrix} q_1 & q_2 \end{matrix}]^T &= [\begin{matrix} x & \theta \end{matrix}]^T
\end{align}
\noindent and the potential energy of the system, given by
\begin{equation}\label{eq36}
\mathcal{F} = \cos \theta
\end{equation}
\noindent the equations of motion can be obtained as
\begin{equation}\label{eq37}
\left[\begin{matrix} 2 & \cos\theta \cr \cos\theta & 1 \end{matrix}\right]\left[\begin{matrix}\ddot x\cr \ddot \theta \end{matrix}\right] - \left[\begin{matrix} \sin\theta\, \dot \theta^2 \cr g \sin \theta \end{matrix}\right] = \left[\begin{matrix} u \cr 0 \end{matrix}\right]
\end{equation}
\noindent which is of the form in (\ref{eq1}). The VHC in (\ref{eq7}) is chosen as 
\begin{equation}\label{eq38}
\rho = x + 1.5 \, \sin \theta = 0
\end{equation}
\noindent which is identical to that considered in \cite{shiriaev2005constructive}. It can be verified that the mass matrix in (\ref{eq37}) and the choice of VHC in (\ref{eq38}) satisfy Assumptions \ref{assum-1} and \ref{assum-2} for $\bar q = (0, \,\, 0)$. For the VHC in (\ref{eq38}) to be stabilizable, Remark \ref{rem1} provides the following condition that needs to be satisfied:
\begin{equation}\label{eq39}
1 - 1.5 \cos^2 \theta \neq 0 \quad \Rightarrow \quad \theta \not= \pm 0.61\,\, {\rm rad}
\end{equation}
\noindent To compare our control design with that presented in \cite{shiriaev2005constructive}, we assume that $\theta \in (-0.61, 0.61)$ such that (\ref{eq39}) is satisfied and the VHC in (\ref{eq38}) is
stabilizable. Simulation results will show that (\ref{eq39}) is indeed satisfied.\

\subsection{Stabilization of VHC and $\mathcal{O}_d$}
The ICPM approach relies on stabilization of both the constraint manifold $\mathcal{C}$, and the orbit $\mathcal{O}_d$ on $\mathcal{C}$. This is a distinctive difference between our approach and the approach in \cite{shiriaev2005constructive} where $\mathcal{O}_d$ is stabilized without stabilizing $\mathcal{C}$, \emph{i.e.}, without enforcing the VHC. To enforce the VHC, we choose the gains $k_p$ and $k_d$ in (\ref{eq11}) as follows:
\begin{align}\label{eq40}
k_p =2, \quad k_d = 1
\end{align}
\noindent We choose the desired orbit $\mathcal{O}_d$ to pass through the point:
\begin{equation}\label{eq40a}
(x,\,\, \theta,\,\, \dot x,\,\, \dot \theta) = (0.0,\,\, 0.0,\,\, -0.675,\,\, 0.450)
\end{equation}
\noindent which is approximately the desired orbit in \cite{shiriaev2005constructive} - see Fig.2 therein. To stabilize $\mathcal{O}_d$, we define the Poinacar\'e section
\begin{align}\label{eq41}
\Sigma = \{x \in \mathbb{Q}^2 \times R^2 \,:\, \theta = 0, \,\, \dot \theta \geq 0\}
\end{align}
\noindent The states of the system on $\Sigma$ are
\begin{equation*}
z = [\,\begin{matrix} x & \dot x & \dot \theta \end{matrix}\,]^T
\end{equation*}
\noindent Since $z^*$ lies on $\mathcal{O}_d$, using (\ref{eq40a}) and (\ref{eq41}) we get
\begin{equation*}
z^* = [\begin{matrix} 0.0 & -0.675 & 0.450\end{matrix}]^T
\end{equation*}
\noindent The matrices $\mathcal{A}$ and $\mathcal{B}$ in (\ref{eq30}) and (\ref{eq31}) are obtained as
\begin{align*}
\mathcal{A} = \left[\begin{array}{rrr}0.115 &  0.435 & 0.600 \cr  -0.510 & -0.640 & -2.465 \cr -0.145 &  0.215 &1.325 \end{array}\right], \quad
\mathcal{B} = \left[\begin{array}{r}-0.06 \cr  1.80 \cr  -1.09 \end{array}\right]
\end{align*}
\noindent It can be verified that the eigenvalues of $\mathcal{A}$ do not lie inside the unit circle but the pair $\{\mathcal{A}, \mathcal{B}\}$ is controllable and satisfy Proposition \ref{prop1}. Using LQR design, the gain matrix $\mathcal{K}$ in (\ref{eq29}) was obtained as
\begin{align}\label{eq42}
\mathcal{K} &= [\begin{array}{rrr}0.163&    0.288&    1.198 \end{array}]
\end{align}
\noindent The eigenvalues of $(\mathcal{A} + \mathcal{B}\mathcal{K})$ are located at $0.13$ and $-0.06 \pm 0.48 i$; this implies that the impulsive feedback exponentially stabilizes the desired orbit
$\mathcal{O}_d$.\

\subsection{Simulation Results}
\begin{figure}[b!]
\centering
\psfrag{A}[][]{\small{time (s)}}
\psfrag{B}[][]{\small{$\rho$}}
\psfrag{C}[][]{\small{$\dot\theta$ vs $\theta$}}
\includegraphics[width=0.90\hsize]{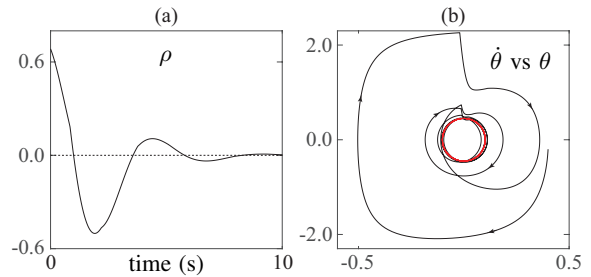}
\caption{Orbital stabilization for the cart-pendulum system; the initial conditions were taken from \cite{shiriaev2005constructive}.}
\label{Fig5}
\end{figure}
The initial configuration of the system is taken from \cite{shiriaev2005constructive}:
\begin{equation*}
[\begin{matrix}x &\theta &\dot x  &\dot\theta\end{matrix}] = [\begin{matrix} 0.1 & 0.4 & -0.1& -0.2 \end{matrix}]
\end{equation*}
\noindent For the controller gains in (\ref{eq40}) and (\ref{eq42}), simulation results for the ICPM are shown in Fig.\ref{Fig5}; $\rho$ is plotted with time in Fig.\ref{Fig5} (a) and the phase portrait of the pendulum is shown in Fig.\ref{Fig5} (b). It can be seen from Fig.\ref{Fig5} (a) that the continuous controller $u_c$ in (\ref{eq11}) enforces the VHC in (\ref{eq38}). To stabilize $\mathcal{O}_d$, the impulsive controller in (\ref{eq29}) is implemented using the high-gain feedback in (\ref{eq34}) with $\Lambda =1$ and $\mu = 0.005$. It can be seen from the phase portrait in Fig.\ref{Fig5} (b) that the pendulum trajectory converges exponentially to $\mathcal{O}_d$, shown in red. The effect of discrete impulsive feedback can be seen in Fig.\ref{Fig5} (b) where $\dot\theta$ jumps when trajectories cross the Poincar\'e section $\Sigma$ defined in (\ref{eq41}). The system trajectories reach a close neighborhood of $\mathcal{O}_d$ in approximately $10$ sec; this is comparable to the results in \cite{shiriaev2005constructive}.\

\begin{figure}[t!]
\centering
\psfrag{A}[][]{\small{time (s)}}
\psfrag{B}[][]{\small{$\rho$}}
\psfrag{C}[][]{\small{$\dot\theta$ vs $\theta$}}
\includegraphics[width=0.90\hsize]{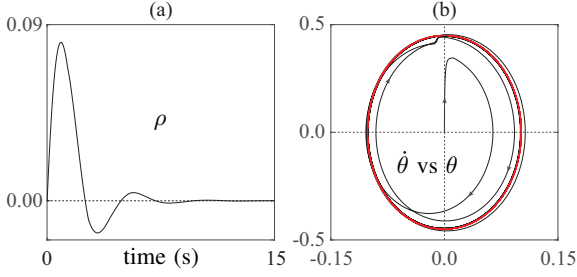}
\caption{Orbital stabilization for the cart-pendulum system for the initial conditions in (\ref{eq42a}).}
\label{Fig6}
\end{figure}

We now consider the following initial condition that lies far away from $\mathcal{O}_d$:
\begin{equation} \label{eq42a}
[\begin{matrix}x &\theta &\dot x  &\dot\theta\end{matrix}] = [\begin{matrix} 0.0 & 0.0 & 0.0& 0.0 \end{matrix}]
\end{equation}
\noindent We used the same controller gains as that used in the previous simulation. It can be seen from the results shown in Fig.\ref{Fig6} that $\mathcal{O}_d$ is stabilized. For the initial conditions in (\ref{eq42a}), the control design in \cite{shiriaev2005constructive} fails to converge the pendulum trajectory to $\mathcal{O}_d$; this implies that $\mathcal{O}_d$ has a larger region of attraction with the ICPM approach than with the control design in \cite{shiriaev2005constructive}. To demonstrate the generality of the ICPM approach, we consider the three DOF tiptoebot, which is presented next.
\section{Illustrative Example - The Tiptoebot} \label{sec-5}
\begin{figure}[b!]
\centering
\psfrag{A}[][]{\footnotesize{$\ell_1$}}
\psfrag{B}[][]{\footnotesize{$\ell_2$}}
\psfrag{C}[][]{\footnotesize{$\ell_3$}}
\psfrag{D}[][]{\footnotesize{$\theta_1$}}
\psfrag{E}[][]{\footnotesize{$\theta_2$}}
\psfrag{F}[][]{\footnotesize{$\theta_3$}}
\psfrag{M}[][]{\footnotesize{$\tau_2$}}
\psfrag{N}[][]{\footnotesize{$\tau_3$}}
\psfrag{X}[][]{\footnotesize{$x$}}
\psfrag{Y}[][]{\footnotesize{$y$}}
\psfrag{G}[][]{\footnotesize{$g$}}
\includegraphics[width=0.30\hsize]{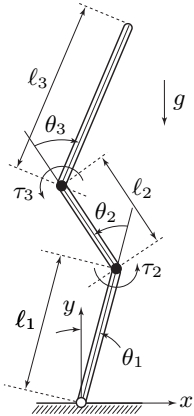}
\caption{The three-link underactuated tiptoebot.}
\label{Fig1}
\end{figure}

\subsection{System Description}\label{sec5-1}
Consider the three DOF tiptoebot \cite{kant2019estimation} shown in Fig.\ref{Fig1}. The tiptoebot is a human-like underactuated system with one passive joint; the three links are analogous to the lower leg, upper leg and torso. The knee joint connecting the upper and lower legs, and the hip joint connecting the torso and upper leg are active; the torques applied by the actuators in these joints are assumed to be positive in the counter-clockwise direction and are denoted by $\tau_2$ and $\tau_3$. The toe provides a point of support and is a passive revolute joint. The joint angles of the links, $\theta_1$, $\theta_2$ and $\theta_3$, are measured positive in the counter-clockwise direction; $\theta_1$ is measured relative to the $y$-axis and $\theta_2$ and $\theta_3$ are measured relative to the first and second links. Using the following definition for the joint angles and control inputs
\begin{align}\label{eq43}
q_1^T = [\begin{array}{cc} \theta_2 &\theta_3 \end{array} ]^T, \quad q_2= \theta_1, \quad u = [\tau_2 \,\,\,\tau_3 ]^T
\end{align}
\noindent the dynamics of the tiptoebot can be expressed in the form given in (\ref {eq1}), where the components of the mass matrix and the potential energy are:
\begin{equation}\label{eq44}
\begin{aligned}
M_{11} &= \begin{bmatrix} \alpha_2 \!+\! \alpha_3 \!+\! 2\alpha_5\cos \theta_3 \,&\, \alpha_3 \!+\! \alpha_5\cos \theta_3\\ \alpha_3 \!+\! \alpha_5\cos \theta_3 & \alpha_3 \end{bmatrix}\cr
M_{12} &= \begin{bmatrix}\alpha_2 \!+\! \alpha_3 \!+\! \alpha_4\cos \theta_2 \!+\! 2\alpha_5\cos\theta_3 \!+\! \alpha_6\cos(\theta_2\!+\!\theta_3) \\ \alpha_3 \!+\! \alpha_5\cos\theta_3 \!+\! 
\alpha_6\cos(\theta_2\!+\!\theta_3)\end{bmatrix} \cr
M_{22} &= \alpha_1 \!+\! \alpha_2 \!+\! \alpha_3 \cr
&+ 2\left[\alpha_4\cos \theta_2 + \alpha_5\cos\theta_3 \!+\! \alpha_6\cos(\theta_2\!+\!\theta_3)\right] \cr
\mathcal{F} =\,\, &\beta_1\cos\theta_1 + \beta_2\cos(\theta_1+\theta_2) +\beta_3\cos(\theta_1+\theta_2+\theta_3)
\end{aligned}
\end{equation}
\noindent where $\alpha_i$, $i = 1, 2, \cdots, 6$, and $\beta_i$, $i = 1, 2, 3$ are lumped physical parameters; their values are given in Table \ref{Tab1}. It can be verified that Assumption \ref{assum-1} is satisfied for $\bar q = (0\,\,\, 0\,\,\, 0)^T$.
\begin{center}
\captionof{table}{Tiptoebot lumped parameters in SI units}
\label{Tab1}
\begin{tabular}{|p{0.5cm}|p{1cm}|p{0.5cm}|p{1cm}|p{0.5cm}|p{1cm}|}
\hline
$\alpha_1$ & $0.386$ & $\alpha_4$  &  $0.065$ & $\beta_1$ & $4.307$\\
\hline
$\alpha_2$ & $0.217$ & $\alpha_5$  &  $0.054$ & $\beta_2$ & $1.102$\\
\hline
$\alpha_3$ & $0.247$ & $\alpha_6$  &  $0.104$ & $\beta_3$ & $1.764$\\
\hline
\end{tabular}
\end{center}
\subsection{Imposing VHC and Selection of $\mathcal{O}_d$}\label{sec5-2}
The VHC in (\ref{eq7}) is chosen as
\begin{align}\label{eq45}
\rho = \left[\begin{matrix}\rho_1 \cr \rho_2\end{matrix}\right] = \left[\begin{matrix}\theta_2 - A_1 \theta_1 \cr \theta_3 - A_2\theta_1\end{matrix}\right] = \left[\begin{matrix}0 \cr 0\end{matrix}\right]
\end{align}
\noindent where $A_1 = -2$ and $A_2 = 0.1$. It can be verified that the VHC in (\ref{eq45}) satisfies Assumption \ref{assum-2} for $\bar q = (0\,\,\, 0\,\,\, 0)^T$; also, it is stabilizable as it satisfies the condition in Remark \ref{rem1}. To enforce the VHC, the gain matrices in (\ref{eq11}) were chosen as
\begin{align}\label{eq46}
k_p = \begin{bmatrix}1.0 & 0.0\\0.0 & 1.0\end{bmatrix}, \quad k_d = \begin{bmatrix}0.1 & 0.0 \\0.0 & 0.1\end{bmatrix}
\end{align}
\noindent The phase portrait of the zero dynamics in (\ref{eq14}) is shown in Fig.\ref{Fig2}. It can be seen that the equilibrium $(\theta_1, \dot \theta_1) = (0, 0)$ is a center, surrounded by a dense set of closed orbits. We choose the desired orbit $\mathcal{O}_d$ to be the one that passes through $(\theta_1, \dot \theta_1) = (0.0, 3.0)$.

\subsection{Stabilization of $\mathcal{O}_d$}\label{sec5-3}
\begin{figure}[b!]
\centering
\psfrag{A}[][]{\small{$\theta_1$ (rad)}}
\psfrag{B}[][]{\small{$\dot\theta_1$ (rad/s)}}
\psfrag{C}[][]{\small{$\mathcal{O}_d$}}
\includegraphics[width=0.64\hsize]{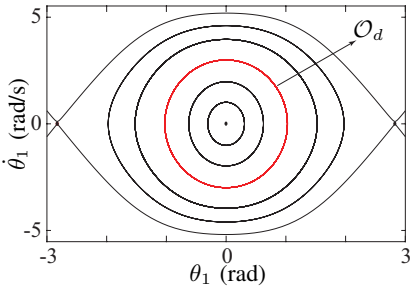}
\caption{Phase portrait of tiptoebot zero dynamics.}
\label{Fig2}
\end{figure}
The desired orbit $\mathcal{O}_d$, shown in red in Fig.\ref{Fig2}, is symmetric about $\theta_1 = 0$ and without loss of generality we define the Poincar\'e section of $\mathcal{O}_d$ as follows
\begin{align}\label{eq47}
\Sigma = \{x \in \mathbb{Q}^3 \times R^3 \,:\, \theta_1 = 0, \,\, \dot \theta_1 \geq 0\}
\end{align}
\noindent The states on $\Sigma$ are
\begin{equation*}
z = [\begin{matrix} \theta_2 & \theta_3 & \dot \theta_1  & \dot \theta_2 & \dot \theta_3\end{matrix}]^T
\end{equation*}
\noindent The fixed point $z=z^*$ lies on $\mathcal{O}_d$ and satisfies the VHC relationship $\rho = \dot \rho =0$. Substituting $(\theta_1, \dot \theta_1) = (0.0, 3.0)$ in (\ref{eq45}) and its derivative gives
\begin{equation}\label{eq48}
z^* = [\begin{matrix} 0.0 & 0.0 & 3.0  & -6.0 & 0.3\end{matrix}]^T
\end{equation}
\setlength\arraycolsep{2.5pt}
\noindent The matrices $\mathcal{A}$ and $\mathcal{B}$ in (\ref{eq30}) and (\ref{eq31}) were obtained as
\begin{equation*}
\begin{aligned}
\mathcal{A} &= \left[\begin{array}{rrrrr}-0.380 &   -0.080 & 1.530 & 0.800 & 0.050 \cr  0.000 &   -0.460 & -0.080 & -0.003  &  0.730 \cr
1.230 &  1.890 & 6.120 & 2.770 & 4.050 \cr -3.210 & -3.770 & -13.360 & -6.090 & -8.100 \cr 0.120 &   -0.560 &  0.670 & 0.280 & 0.100\end{array}\right] \\
\mathcal{B} &= \left[\begin{array}{rrrrr}1.525 &  -3.700 &  -17.700 &   34.325 &   0.875 \cr  
4.875 &  -8.650 &   22.650 &  -43.850 &  -0.325 \end{array}\right]^T
\end{aligned}
\end{equation*}
\noindent The eigenvalues of $\mathcal{A}$ do not lie inside the unit circle but the pair $\{\mathcal{A}, \mathcal{B}\}$ is stabilizable and satisfy Proposition \ref{prop1}. Using LQR, the gain matrix $\mathcal{K}$ in (\ref{eq29}) is obtained as
\begin{align}\label{eq49}
\mathcal{K} &= \left[\begin{array}{rrrrr}0.028 &  0.024 & 0.197 &  0.094 & 0.138 \cr  
-0.034 &  -0.051 &   0.116 &  -0.049 &  -0.055 \end{array}\right]
\end{align}
\noindent The eigenvalues of $(\mathcal{A} + \mathcal{B}\mathcal{K})$ are located at $0.14$, $-0.47 \pm 0.73 i$ and $-0.12 \pm 0.56 i$; this implies that $\mathcal{O}_d$ is exponentially stable.

\subsection{Simulation Results}
\begin{figure}[t!]
\centering
\psfrag{A}[][]{\small{time (s)}}
\psfrag{B}[][]{\small{$\rho_1$}}
\psfrag{C}[][]{\small{$\dot\rho_1$}}
\psfrag{D}[][]{\small{$\rho_2$}}
\psfrag{E}[][]{\small{$\dot\rho_2$}}
\psfrag{M}[][]{\small{$\|e(k)\|_2$}}
\psfrag{F}[][]{\small{$k$}}
\includegraphics[width=0.99\hsize]{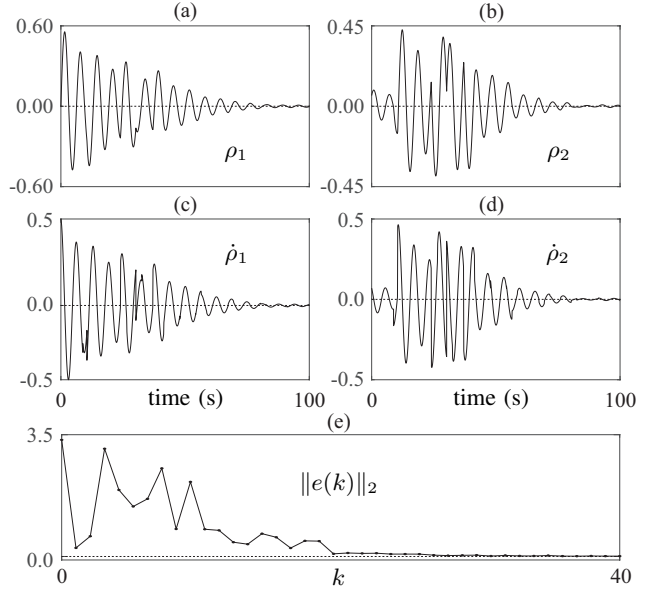}
\caption{Orbital stabilization for the tiptoebot using ICPM.}
\label{Fig3}
\end{figure}

The initial configuration of the tiptoebot is taken as
\begin{equation*}
[\begin{matrix}\theta_1 &\theta_2 &\theta_3 &\dot\theta_1  &\dot\theta_2 &\dot \theta_3\end{matrix}] = 
[\begin{matrix} -0.1 & 0.2 & 0.05 & 3.3 & -6.0 & 0.4 \end{matrix}]
\end{equation*}
For the controller gains in (\ref{eq46}) and (\ref{eq49}), simulation results of the ICPM approach are shown in Fig.\ref{Fig3}. The plots of $\rho_1$, $\rho_2$, $\dot\rho_1$ and $\dot\rho_2$ with time are shown in Figs.\ref{Fig3} (a)-(d); it can be seen that the continuous controller $u_c$ in (\ref{eq11}) enforces the VHC in (\ref{eq45}) by stabilizing the constraint manifold $\mathcal{C}$. To stabilize the desired orbit $\mathcal{O}_d$, the impulsive controller in (\ref{eq29}) is implemented using the high-gain feedback in (\ref{eq34}); $\Lambda $ was chosen to be an identity matrix and $\mu$ was chosen as $0.0001$. To show the convergence of system trajectories to $\mathcal{O}_d$, $\| e(k) \|_2$ is plotted with respect to $k$ in Fig.\ref{Fig3} (e). It can be seen that for large values of $k$, $\| e(k) \|_2 \rightarrow 0$; this implies that $\mathcal{O}_d$ is exponentially stable.\

\section{Conclusion}\label{sec7}
Repetitive motion in underactuated systems are typically designed using VHCs. A VHC results in a family of periodic orbits and stabilization of an orbit is an important problem in applications such as legged locomotion. A hybrid control design was presented to stabilize a VHC-generated periodic orbit for underactuated system with one passive DOF; a continuous controller was used to enforce the VHC and impulsive inputs were periodically applied on a Poincar\'e section to stabilize the desired orbit. These impulsive inputs alter the Poincar\'e map and this impulse controlled Poincar\'e map (ICPM) is described by a discrete time-invariant linear system. The problem of orbital stabilization problem is thus simplified to stabilization of the fixed point of the ICPM. The controllability of the system can be easily verified and the control design can be easily carried out using standard techniques such as pole-placement and LQR. The identification of the linear system and computation of the controller gains are performed off-line. The complexity and computational cost of the ICPM approach is less than existing methods in the literature as it eliminates the need for on-line solution of a periodic Ricatti equation. The ICPM approach is demonstrated using the standard cart-pendulum system; its applicability to higher-dimensional systems is demonstrated using the three-DOF tiptoebot. Future work will focus on gait stabilization of legged robots undergoing ground-foot impacts and experimental validation.\

\balance
\bibliography{ref}
\bibliographystyle{plain}  

\end{document}